# FACULTY ATTITUDES TOWARDS INTEGRATING TECHNOLOGY AND INNOVATION

Colleen Marzilli[1], Julie Delello[2], Shelly Marmion[3], Rochell McWhorter[4], Paul Roberts[5] and T. Scott Marzilli[6]

[1]College of Nursing, The University of Texas at Tyler, Tyler, Texas, USA
[2]School of Education, The University of Texas at Tyler, Tyler, Texas, USA
[3]Department of Psychology and Counseling, The University of Texas at Tyler, Tyler, Texas, USA
[4]Department of Human Resource Development and Technology, The University of Texas at Tyler, Tyler, Texas, USA
[5]College of Business and Technology, The University of Texas at Tyler, Tyler, Texas, USA
[6]Academic Innovation and Student Success, The University of Texas at Tyler, Tyler, Texas, USA

## ABSTRACT

*Technological innovation is an important aspect of teaching and learning in the 21st century. This article examines faculty attitudes toward technology use in the classroom at one regional public university in the United States. Building on a faculty-led initiative to develop a Community of Practice for improving education, this study used a mixed-method approach of a faculty-developed, electronic survey to assess this topic. Findings from 72 faculty members revealed an overall positive stance toward technology in the classroom and the average faculty member utilized about six technology tools in their courses. The opportunities, barriers and future uses for technologies in the higher education classroom emerged from the open-ended questions on the survey. One finding of particular concern is that faculty are fearful that technology causes a loss of the humanistic perspective in education. The university is redesigning ten of its most popular courses to increase flexibility, accessibility and student success.*

## 1. INTRODUCTION

Traditional students have spent their entire lives surrounded by digital technologies [1][2]. Through their use of cellphones, smartphones, tablet computers and laptops, college students are arriving in higher education classrooms more technologically linked and socially connected than ever before [3]. These portable technologies with online connectivity challenge educators to meet students in the technological world where they now live [4][5]. In a recent study by the Pew Research Center, 60% of the experts and stakeholders surveyed predicted innovative shifts would occur in higher education by the year 2020 [6]. These innovative changes include "cloud-based computing, digital textbooks, mobile connectivity, high-quality streaming video and "just-in-time" information gathering" (p. 2).

From flipped classrooms to massive open online courses (MOOCs), eLearning is creating a notable transformation in higher education. As the paradigm shifts from traditional teaching methods to technology-enabled learning, it is essential that instructors be well prepared to utilize new technologies to meet the needs of all students. According to Prensky [7], "To create effective 21st century learning… students need to be allowed to do new things, in new ways, and get a different, and better, education because of the technology" (p. 1). However, the teaching model in higher education is inconsistent with the technological needs of these millennial learners [8][9]. Davis [10] makes the case that the fundamental issue is "grounded in the debate between those who want increased technology in the classroom and those who are concerned that increased technology will prevent or limit what they see as effective instruction" (p. 1).

It is vital for instructors to recognize both the opportunities and challenges required to meet the demands of the 21st century learner. However, while innovative practices, new technologies, and eLearning proliferate in higher education, methods to precisely define and measure teacher effectiveness are still under considerable debate [11][12]. Furthermore, stakeholders must comprehend the attributes of effective implementation of new technologies in the higher education environment including faculty members' level of readiness for implementation as well as their fears, preferences, teaching styles, and passions [13][14].

This article describes the creation of a Faculty Learning Community (FLC) at one public university to examine the innovative use of technologies on campus. Further, the purpose of this mixed-methods case study is to understand the opportunities and challenges faculty members face as they strive to keep pace with the development of online and hybrid courses and the pervasive spread of new and innovative classroom technologies.

## 1.1. Review of the Literature

### 1.1.1. Innovation and Technology

In his 2011 State of the Union address, President Obama called for a nationwide emphasis on innovation stating, "We need to out-innovate, out-educate, and out-build the rest of the world… the first step in winning the future is encouraging American innovation" (para. 22) [15]. The literature revealed that true innovation is not about a newfangled invention. It is about individuals and how they use ideas in new ways. According to Lichtman, innovation is not technology or flipped classrooms but rather how education prepares "students for their futures not for our past" (p. 8) [16].

To create an innovative classroom, The New Media Consortium [17] conveyed that teaching content is no longer sufficient by stating: "Students expect learning that matters; learning connected in timely ways to the real world; learning that engages their interests; and learning experiences that see them as entire persons, not just consumers of content" (p. 2). With this statement in mind, the question for faculty members is how do they balance the required content with future innovations?

According to one researcher [18], in order to integrate innovation into the curriculum and pedagogy, higher education needs to emphasize the Five I's: Imagination, Inquiry, Invention, Implementation, and Initiative into a system that often discourages creative thinking and risk-taking. In accordance with The National Educational Technology Standards for Teachers (NETS-T), the performance expectation is that all educators (K-16) will "promote, support, and model creative and innovative thinking and inventiveness" (para. 2) [19]. This is often supported with the inclusion of technological innovations in the curriculum.

### 1.1.2. Technology Transformation

As the demand for mobile technology and personalized learning propagates, a clear transformation in the use of technology must occur. It seems that educators have more choices to engage students than ever before. Steven Mintz, Executive Director of The University of Texas System's Institute for Transformational Learning pointed out that the future of higher education will include "much higher levels of interaction through collaborative learning, as well as animations, educational gaming, immersive-learning environments, and hands-on simulations" (para 12) [20]. However, the most standard uses of technologies in classrooms rely heavily on lectures and learner management systems (LMS). On average, according to the Faculty Survey of Student Engagement, faculty members teaching courses in Biological Sciences, Engineering, and Physical Sciences allocated more than half of their class time to lecturing whereas faculty members in Education committed less than one quarter, indicating variation across disciplines in the use of innovative teaching techniques versus traditional lecturing and low-tech teaching strategies [21]. Green indicated in the 21st National Survey of Computing and Information Technology in American Higher Education study that learning management systems were the primary platform favored by faculty members [22]. The Educause Center for Applied Research (ECAR) reported that 90% of institutions of higher education currently use a LMS to deliver content to students [23]. Yet, according to the Faculty Survey of Student Engagement (2009) much of this use is predominantly for faculty members to distribute resources, grades, or syllabus information to students. Additionally, most faculty members still elect to use asynchronous discussions (LMS, email) rather than synchronous discussions (phone, videoconferencing) with students [24][25].

Web 2.0 utilizes Internet based platforms such as social media sites for online collaboration. While the majority of higher education students participate in social networking in their personal lives; currently,there is a growing consensus that social media platforms can be used for innovative teaching and learning to engage students in a meaningful way [26]. The Babson Survey Research Group, in collaboration with New Marketing Labs and Pearson Learning Solutions, reported that 80% of higher education faculty respondents used social media in the classroom and over half of that use is for instructional purposes [27]. These groups collected data from 1,920 faculty members highlighting the fact that "virtually all higher education teaching faculty members are aware of the major social media sites" (p. 3) [28] while in another study, 44.7% of faculty stated that they never use social media for formal learning [29]. Many faculty members attributed the primary reasons for the lack of adoption of social media to student privacy concerns and plagiarism issues [28].

In the study *Going the Distance: Online Education in the United States*, over six million students reported that they had taken at least one online course during the Fall of 2010 [30]. However, in a recent Gallup poll, only one in five faculty members perceived online teaching to be equivalent to face-to-face courses [31]. Furthermore, increased access to higher education through digital technologies has been perceived as a threat by instructors who prefer the traditional mode of course delivery. The New Media Consortium report noted that faculty members who teach in one way may be hesitant to invest the time to learn new methods or may lack the budget for needed support [32]. Additionally, only 6.3% of faculty members reported in the *Digital Faculty Study* that their institution rewarded faculty members' contributions to digital pedagogy while 65% of instructors reported an increased workload [29]. If true innovation is to occur in higher education, it seems reasonable that a technology transformation must also occur. Faculty members' readiness and willingness are crucial to the success of such a transformation.

### 1.1.3. Theory for Innovation

The conceptual framework for this study and the FLC is built upon the notion of a community of practice (CoP). Communities of practice have been described by [33] as "groups of people who share a concern or a passion for something they do and learn how to do it better as they interact regularly" (p. 98). Members of a CoP, according to Wenger, are practitioners who share both resources and experiences with one another. The researcher stated: "Technology platforms are often described in terms of features, but in order to really evaluate candidates for a technology platform, it is useful to start with the success factors of communities of practice that can be affected by technology" (p. 4) [34]. According to the United States Department of Education [35], building CoP among teachers and leaders is widely considered to be a promising strategy for sustained, substantive institutional improvement. Real innovation, as stated by [36], is about building a community of professionals while providing a forum for innovations.

### 1.1.4. Context

In 2012, the faculty at one regional university in the southwest United States developed a FLC on Teaching Innovations as part of the CoP model. The purpose of the FLC was to investigate, encourage, model, and share innovations in teaching, particularly focusing on the use of technology to improve classroom instruction.

The goals of the FLC were to learn about the use of technology to enhance teaching and to share that information, but also to try out innovative techniques in classes in order to see the impact on teaching and learning. The FLC focused on research, improved teaching, increased retention rates, supplemental instruction, and the future development of a Center for Teaching Excellence and Innovation on the campus. In order to gather information for improving the instructional outcomes at the university, it was important for the FLC to conduct an assessment of the current faculty members' technology practices.

## 2. METHODOLOGY

Within the framework of the FLC and the CoP, the researchers used the mixed-method paradigm to guide the case study. According to Plano Clark and Creswell, mixed-method research uses both quantitative and qualitative approaches in combination in order to provide a "better understanding of research problems than either approach alone" (p. 5) [37].

This study aimed to consider the perceptions of faculty members in regards to their levels of technology readiness to incorporate innovative technologies for student learning. To accomplish the purpose of this study, the following research questions will be addressed:

1. What are faculty members' attitudes toward technology in the higher education classroom and how does that relate to their reported skills and usage?
2. What opportunities are identified by faculty members for using technologies in the higher education classroom?
3. What barriers are identified by for using technologies in the higher education classroom?
4. How do faculty members perceive the future of using technology in the higher education classroom?

## 2.1. Procedure

The assessment consisted of developing a mixed-methods survey. A convenience sample of 72 full-time faculty members representing a 25% response rate from across five colleges in 32 departments participated in this descriptive mixed-methods study. Prior to data collection, Institutional Review Board approval was obtained. Inclusion criteria included full-time faculty members as of September 1, 2012.

Guidelines for the protection of human subjects were followed. An email invitation described the purpose of the study, instructions for completing the survey, risks and benefits of participation, the option to withdraw from the study, and included a link to the electronic survey.

## 2.2. Instrumentation

A mixed-methods survey was created assessing faculty members' self-reported level of skill and usage of technology in the classroom, attitudes towards its use and usefulness, perceptions of the opportunities afforded by its use, barriers to usage, and predictions of the future of technology in the classroom. The survey was examined by five FLC members with appropriate expertise to assess for content validity and clarity, and clarification regarding any ambiguous and redundant information was made. The final survey was agreed upon by all FLC team members and delivered via the electronic survey system.

The final survey included 40 categorical, quantitative questions and six qualitative questions. For each of the items requiring a narrative response, additional open-ended questions were asked including the following: 1) What are some of the ways you have used technology in your coursework?; 2) What technologies do you find the most useful to teaching?; 3) What technologies do you find the least useful to teaching?; 4) Are there problems with technology in classes that have interfered with teaching and learning?; 5) What barriers do you experience when using technology for educational purposes?; and 6) What do you think technology will look like in the future?" The open-ended questions were used to obtain more comprehensive information as well as to better understand the attitudes of faculty members towards technology.

## 3. DATA ANALYSIS

The mixed-method approach to collecting data provides the researchers with an opportunity to determine both quantitative and qualitative data to ensure richness of the data. Both quantitative questions and the qualitative questions were included to provide a more comprehensive understanding of the faculty members' attitudes towards technology in a way that would not be possible using only a quantitative or a qualitative approach [37].

## 3.1. Data Analysis of Survey Item Responses

Descriptive statistics were used to measure the central tendency and distribution of responses for quantifiable survey responses. Composite variables were formed from sets of related responses. Using SPSS 20, correlations between composite variables were performed, as well as t-test comparisons of groups varying in their preferred amount of technology using the composite variables.

## 3.2. Data Analysis of Open-ended Survey Responses

The researchers employed a survey to obtain qualitative data through open-ended questions. The data for the six open-ended questions were divided into three categories including opportunities, barriers, and the future. The data was hand coded based on recurring extracts within each category. Researchers separately coded the data, and then the results were compared between them for reliability. The researchers agreed upon consistent themes using the similarity principle [38].

## 4. FINDINGS

### 4.1. Faculty Attitudes

The survey included four items which addressed the first research question: *What are faculty members' attitudes toward technology in the higher education classroom and how does that relate to their reported skills and usage?*

The first item, *Preferred Amount*, asked respondents to choose a statement that best described their preference with respect to the amount of technology incorporated into teaching. "Limited" was chosen by only 7.1%, "Moderate" was chosen by 50%, while 35.7% and 7.1% chose "Extensive" and "Exclusive," respectively.

The other three items which measured faculty members' attitudes asked them to indicate their agreement or not with the following three positive statements regarding technology use in the classroom. Technology...(1) creates excitement and enhances learning, (2) improves my teaching, and (3) makes teaching more convenient. Responses were summed over the three items, producing a continuous measure of the respondents' attitude toward technology in the classroom. The range of possible scores was from 3 to 9, with 9 being agreement with all three positive statements. The mean response was 7.9 (SD = 1.46), again indicating an overall positive stance to the use of technology in the classroom.

*Technology skills* were assessed by asking respondents to describe their skill (No skill, Fairly skilled, or Expert) in using the university website, library website, Blackboard, Excel, Word, and PowerPoint, as well as with computer maintenance, surfing the web, and evaluating the credibility of online information. Responses across these items were summed for a measure of overall skill with these standard technologies. The range of possible scores was from 3 to 27, and the mean for this faculty group was 21.6 (SD = 3.58).

*Course Use* was measured by asking respondents to indicate which of 17 technologies they employed in courses, ranging from word-processing software to simulations and video creation software. The number of differing technologies checked by the respondent was determined, and could vary from 0 to 17. The mean number of technologies reportedly used by respondents in their courses was 5.67 (SD = 2.47).

*Technology Usage Frequency* was measured by asking respondents to indicate the frequency (Never, Infrequently, Frequently) with which they used twenty-one different technologies ranging from text messaging to virtual worlds and video editing. A composite score was created by summing across all 21 technologies. The range of possible scores was from 21 to 63, and the mean for this faculty group was 42.5 (SD = 6.23).

As might be expected, the skill and usage variables were all highly correlated, with the strongest relationship between Course Use and Usage Frequency, r (67) = .52, p < .001. *Course Use* and *Skill Level* also significantly correlated, r (70) = .42, p < .001, as did *Usage Frequency* and Skill, r (66) = .37, p = .003. Of those three variables, only Class Use was predicted by the respondent's overall attitude toward classroom use, r (70) = .41, p < .001.

When the variable *Preferred Amount* was reformulated into two categories, combining those with preferences for Limited and Moderate technology use in the classroom into one category and those with Extensive and Exclusive preferences into a second category, as expected, the latter group reported significantly greater technology Skill, t (67) = -2.59, p = .012, Course Use, t (68) = -4.04, p < .001, and Usage Frequency, t (63) = -5.41, p < .001.

While we had hoped to make comparisons across disciplines or colleges, low response rates from two colleges made such comparisons untenable.

## 4.2. Opportunities for Faculty

The second research question explored the opportunities for faculty members regarding the utilization of technologies for learning. The research question, *What opportunities are identified by faculty members for using technologies in the higher education classroom?*, was answered by two qualitative questions from the survey. There were 110 responses collected from the survey for the two questions. There were 56 responses to the question, *What are some of the ways you have used technology in your coursework?* These 56 responses were combined with the 54 responses to the question, *What technologies do you find the most useful to teaching?*

When analyzing the responses, eight themes emerged from the data: *Lecture Capture, Grading and Assessment, Asynchronous Forums, Synchronous Forums, Social Media, Learning Management Systems (LMS), Image-based Platforms, and Plagiarism Tools*. These themes, rules for inclusion, and selected faculty members' extracts are highlighted in Table 1 and further explained in the subsequent sections.

The first broad theme termed *Lecture Capture* is supported by extracts from faculty members who remarked that technology provided them the opportunity to capture lectures to support student learning. According to Educause, *Lecture capture* is "an umbrella term describing any technology that allows instructors to record what happens in their classrooms and make it available digitally" (para. 6) [39]. For example, one faculty member said: "I use a digital pen screen in conjunction with OneNote and Collaborate/Tegrity to deliver lectures to students in the classroom and over the Internet." This theme is consistent with the literature as exemplified by research that stated the benefits of lecture capture may include increased student engagement, "appealing to students' interests, offering multiple opportunities to access content and providing opportunities for learners to learn at their own pace" (p. 69) [40].

A second theme that emerged and termed *Grading and Assessment* included remarks from faculty members who reported that technology allowed them the opportunity to assess student knowledge and post student grades online. Faculty members noted that they used a combination of hardware (clickers, smartphones) and online software (Polleverywhere.com) for student assessment. For example, one faculty member stated, "I used Polleverywhere.com in class to do informal class polling using cell phones and texting." According to the website [41], responses are automatically graded and each individual student's score is put into an online report card. Grading was enhanced through the use of LMS technologies such as Blackboard. For instance, one faculty member reported "Most assignments are now submitted and graded in Blackboard" while another remarked "I read papers digitally, using track changes and comments, and then upload the graded papers back into Blackboard."

*Asynchronous technology platforms* was the third theme that emerged from the data. This theme included wikis, blogs, email, and discussion forums. Asynchronous learning forums are a popular means to encourage course discussions when instructors and learners are in different places [42, 43]. For instance, one professor remarked that "I use the discussion board of Blackboard to allow my students to explore the most difficult concepts in each unit." Another faculty member commented, "I plan to incorporate a collaborative experience into my Music Theory class, in which students from [my school] and another institution will create video demonstrations, upload them online, and start a dialogue with each other in a blog format."

The fourth theme that emerged from the open-ended questions on the survey was named *Synchronous Technology Platforms* and incorporated the use of Instant Messaging (IM), Chat, and web conferencing (Skype, Collaborate, GoToMeeting). Rather than using static content, synchronous discussions allow users to communicate in "real-time" through phones, instant messaging (IM), screen-sharing, videoconferencing, and face-to-face discussions with the convenience of distance education [42]. For instance, one faculty member responded that Skype and other video conferencing technologies allowed them "to bring experts and real students to the classroom." A second faculty member reported using Blackboard Collaborate "to interact with students." Echoed in the literature, Weiler stated, "Synchronous efforts turn the overall experience from a self-guided learning process to one where each student feels a part of a larger learning community" (p. 8) [44].

Three popular social media platforms, Facebook, Twitter, and YouTube were reported as most often used to support learning in the fifth theme *Social Media*. One professor noted, "I also use Twitter, YouTube, and Facebook nearly every day." A second faculty member stated, "I incorporate appropriate YouTube videos into lecture." The use of social media, according to research, provides faculty members the opportunity to break the limitation of course management systems, enabling innovative and collaborative interactions and personalized constructive learning [45].

The sixth theme was based upon a *Learning Management System*. A LMS is a software platform for managing curriculum, training materials, and evaluation tools, and all faculty members were provided the LMS platform Blackboard. According to Bradford, Porciello, Balkon and Backus, the benefits of Blackboard to students and the faculty member may include "increased availability, quick feedback, improved communication, tracking, and skill building" (p. 303) [46]. One faculty member used Blackboard to post everything including documents such as notes and homework solutions, grades, and lectures. Another faculty member stated, "Blackboard capabilities are good for sending messages, assigning and collecting homework, [and] disseminating materials." Not surprising, over half of the faculty members in this study (n = 61, 56%) indicated they used the Blackboard platform in their courses.

The seventh theme *Image-based Platforms* included the use of PowerPoint, Softchalk, Prezi, InDesign, Photoshop, Illustrator, Animoto, VoiceThread, Google Art Project, Xtranormal, iMovie, online simulations, videos, and visualizations. PowerPoint (n = 33, 30%) and video technology (n = 35, 32%) were the most preferred methods of showing concepts to students. For instance, one faculty member stated, "My lectures are done with PowerPoint and online animations" while another said, "As a historian, the use of PowerPoint to display images while I lecture is a powerful tool to help students visualize the people, places, and events we are discussing and analyzing." The research supported this theme as exemplified by Heer, Bostock, and Ogievetsky, as they stated "the use of well-designed visual representations can replace cognitive calculations with simple perceptual inferences and improve comprehension, memory, and decision making" (p. 59) [47].

*Plagiarism* was the final theme to emerge from the data. SafeAssign anti-plagiarism software is a free tool integrated into the LMS Blackboard platform. In this case, several faculty members reported using the software to detect plagiarism. For instance, one instructor wrote, "All papers are submitted into SafeAssign" while another professor stated, "SafeAssign and the Respondus Lockdown Browser are nice to have as well."

Table 1. Opportunities for using technologies in the higher education classroom.

| # | Name of theme | Explanation/Rule for Inclusion | Faculty Extracts |
|---|---|---|---|
| 1 | Lecture Capture | Faculty remarked that technology allows lectures to be captured and shared with students. | • I Record all of my lectures with Tegrity.<br>• I deliver lectures to students in the classroom and over the Internet. |
| 2 | Grading and Assessment | Faculty remarked that they technology allows for help in assessment and grading. | • The grade book function in Blackboard is a MUST.<br>• [I] run a three week simulation with Blackboard and pollanywhere.com. |
| 3 | Asynchronous Forums | Faculty remarked that wikis, blogs, and discussions enhanced their courses. | • Discussion board for presentation of online cases and discussions.<br>• Students learn to set up Word Press blogs and use the blogs throughout program as online portfolio and writing outlet. |
| 4 | Synchronous Forums | Faculty remarked that technology Instant Messaging (IM), Chat, and Collaborate enriched their courses. | • Skype or use other video conferencing technologies to bring experts and real students to my classroom.<br>• I encourage my students to use Blackboard IM to contact each other and me. |
| 5 | Social Media | Faculty remarked that social media was used in courses. | • Facebook nearly everyday<br>• Require students set up professional Twitter I.D. |
| 6 | LMS | Faculty remarked that an LMS provided the opportunity for content delivery. | • Student submission of projects in electronic from to share with classmates for Blackboard group-learning.<br>• I always use Blackboard and also regularly access the Internet during courses. |
| 7 | Image-based Platforms | Faculty felt that visual platforms and tools provided the opportunity to present content to students. | • Access to animations and videos that explain complicated concepts.<br>• I rely heavily on PowerPoint to provide images to accompany the lecture materials I discuss. |
| 8 | Plagiarism Tools | Faculty remarked that technology tools supported faculty in reducing plagiarism. | • All papers are submitted into SafeAssign.<br>• SafeAssign to check for plagiarism |

## 4.3. Barriers to Using Technology

The third research stated, *What barriers are identified by using technologies in the higher education classroom?* This research question was answered by the three survey questions asked in an open-ended format. One hundred forty-five total responses were collected from the questions which included 1) *What technologies do you find the least useful to teaching? 2) Are there problems with technology in classes that have interfered with teaching and learning?* 3)

*What barriers do you experience when using technology for educational purposes?* Seven themes emerged from the data regarding barriers for using technologies in the higher education classroom.

The first barrier was titled *Distracted by Technology* and included remarks by faculty that technology reduces student engagement and that students lack the self-discipline to focus on learning (See Table 2). For example, one participant remarked, "Students today are falling in] to the trap of edutainment. An increasing amount of 'average' students *expect* for their attention to be held by flashes and whirls…educational institutions are somewhat responsible for letting this happen by encouraging so-called 'technology literacy' over actual literacy." This theme is buttressed by the literature, "All of [the distractions of technology have] implications for teaching. The cost of classroom multitasking…can be a failure to learn" (para. 62) [48].

The second theme was titled *Lack of Knowledge - Faculty* and included remarks by faculty that they had a lack of knowledge about technology. For instance, one faculty member remarked, "The main barrier to teaching] is trying to stay current on the available technology - how to use each new iteration…the little seemingly slight changes made with something like the user interface in BB…and]trying to figure them out on several different systems." Another wrote, "If you ask students to do [technology], they have to learn it, and often run into issues that are out of my league to address."

*Lack of Knowledge – Students* was the name of the third theme around the barriers for using technology in the higher education classroom. This theme is concerned with faculty members who found technology and related concepts too difficult for students. One faculty member commented that a barrier to learning was, "Access and understanding on the part of students when it comes to how to use the technology." Another said, "There is always a learning curve with new technology. We will be facing them ever more frequently in the future." This theme is inconsistent with the literature regarding students' knowledge of technology [49].

The fourth theme was titled *Insufficient Resources* and included comments from faculty members who did not feel that they have the resources they need to be successful with technology. One remarked that a barrier was "Inadequate support at both the university and provider levels." Also, another faculty member commented that a barrier was, "Lack of faculty time to locate and incorporate technologies--Our time is EXTREMELY packed now, and it is almost easier to continue on the same course, rather than spend the amount of time required to incorporate newer ideas."

*Unreliability of hardware or software platforms* was the fifth theme that emerged from the data. In this theme, the faculty members felt the technology access on campus was unreliable. For instance, one comment included, "service outages have been a bigger problem this semester. Blackboard went down the day of an exam and the remainder of the course had to be rescheduled." Another faculty member remarked, "Classroom computers are inconsistent or unpredictable" while another said a barrier was the "lack of user friendly technology."

A sixth theme found was named *Pressure from Administrators and Students* whereby faculty members were concerned that students had expectations that were unrealistic. For example, one faculty member wrote, "I get email at 2 a.m. wondering why I haven't responded to their 10 p.m. email. They expect immediate turn-around." Another said, "I don't like feeling pressure to be finding a new way to integrate technology in the classroom every semester. And I do feel that pressure--it is pressure to use technology purely for the sake of using it."

The seventh and final theme was titled, *Outdated Platforms/Tools* which referred to faculty remarks that they found the technology tools to be dated or not useful. For instance, one faculty remarked that, "Students (and I) are beginning to regard PowerPoint as "old school". This theme is echoed in the literature by Hurn, "Universities are struggling to keep up with… new technology, with outdated intranet systems and limited research into its application within the higher education sector" (p. 35) [50].

Table 2. Barriers for using technologies in the higher education classroom.

| # | Name of theme | Explanation/Rule for Inclusion | Faculty Extracts |
|---|---|---|---|
| 1 | Distracted by Technology | Faculty remarked that technology reduces student engagement and that students lack the self-discipline to focus on learning. | • Students don't pay attention to lecture because they are drawn to [their] computer screen.<br>• Teacher must compete with Facebook issues in the class. |
| 2 | Lack of Knowledge - Faculty | Faculty remarked that they had a lack of knowledge about technology. | • My age….I am not as comfortable with technology as younger faculty and it takes me longer to learn new technology.<br>• I have the basics [of Blackboard] down but I am sure it has features I don't even know about. |
| 3 | Lack of Knowledge - Students | Faculty found technology and related issues was too difficult for students. | • Some students are not as tech savvy as expected.<br>• Many students do not have the understanding of how to determine validity of internet information. |
| 4 | Insufficient Resources | Faculty did not feel that they have the resources they need to be successful with technology. | • The level of security makes it difficult for faculty and students to try out software.<br>• My classes are all 40+ and there is no computer lab large enough to use for class which would be very helpful. |
| 5 | Unreliability of Hardware or Software Platforms | Faculty felt the technology and access on campus was unreliable. | • The classroom technology is unreliable.<br>• The current library entry page is one of the most unfriendly places in all of the university sites. It is very, very difficult to access articles through our library. |
| 6 | Pressure from Administrators and Students | Faculty felt that students had expectations that were unrealistic. | • Students are so used to immediate feedback that I've noticed they get frustrated when you can't exist in their electronic world all the time.<br>• I do not like the way some officials are increasingly turning to online classes to replace in-class teaching. |
| 7 | Outdated Platforms/Tools | Faculty found the tools to be dated or not useful. | • I rarely use the Elmo in any classroom-Would prefer more Smart Boards.<br>• I would prefer that we move to another platform such as Moodle that is more user-friendly for courses that are specifically online. |

### 4.4. The Future of Technology

The fourth research question concerning the perception of the future of technology in the educational classroom was addressed by the question *What do you think technology in education will look like in the future?* The 45 responses yielded seven themes (see Table 4).

The first theme in the literature, *Increase in Technology Usage* was very prominent among the responses and is supported in the literature [51], and as one faculty member put it "I think there will be more and more technology used in education. Our students have grown up with technology, and are very accustomed to it." Although there was strong consensus concerning the increase in technology, there was also support for traditional instruction.

The second theme was *Face-to-Face Instruction and Physical Materials*. Faculty members still see strong value in a more traditional approach to higher education. One comment that clearly illustrates this is: "I don't want to lose face-to-face instruction and interaction. I believe conversation and dialogue is a strong form of inquiry." The faculty members predicted the future will bring a technology enriched classroom.

The third theme was *Hybrid Formats, Online Learning, and Real-Time Interaction*. The perspectives of faculty members on this theme are somewhat varied as they range from the addition of technology in the classroom to a move to technology playing a major role in all classes. The following quote demonstrates how technology may change the role of the professor:

All education will be hybrid in the future. The professors' role will be the orchestration of and the focus on needed concepts and information. Students need to be taught how to use and manage the overwhelming amounts information available to them. Teachers will need to be platform agonistic.

*Technology Utilized to Prepare Students for Workplace* was the fourth theme. The ability of students to use technology is seen as a major learning objective in many areas and faculty members recognize the need for students to be proficient in the use of career specific tools which have an ever increasing integration to technology in one form or another. There is also the desire for universities to produce a technologically literate workforce.

The fifth theme was *Online Learning*. Although this was a clear theme, faculty members were mixed on their view of online learning in the future as is illustrated by the following: "Seamless and paperless. No books unless on line, all documentation on line, most lectures on line, choices to read, listen to these lectures. No delays in transmission of information not the annoying delay on the other campuses." Another faculty responded with the following quote:

While it definitely has its perks, I fear those perhaps not familiar with the everyday in-class experience are pushing for it to take a greater role than is effective in teaching, and I fear this domination of online teaching is both a major threat to the kinds of thinking students are supposed to acquire in a university setting but also the path that technology in education will take if there isn't a greater flexibility and understanding of different teaching approaches or an acknowledgement that some non-technology-dependent methods are still useful and not 'archaic' or 'old-fashioned.'

*The New Model of Education* was the sixth theme found in the literature. This may be one of the most interesting themes to emerge from this study. Universities have long demonstrated a resistance to change and greatly lagged industry in the implementation and integration of technology. Many faculty members are predicting a future of higher education in which a quantum leap occurs and the traditional classroom is replaced by low-cost online options. This is supported in the literature [48, 52], and it is also supported in the data by the faculty members' quote below:

It may go to an extreme if what Bill Gates gets what he wants--get rid of university faculty member and hire them back as teaching assistants after putting all college education online. What faculty would do is to assist student collaborative learning from projects, or the so-called project based learning. All the humanistic values of higher education would be gone. Instead, college education would be controlled by the few who have money to buy and sell technology. Look at what's happening to the library databases. If we got rid all the books and relied on the databases only, what would happen? It has already happened. Our access to knowledge and information is being controlled by the database. And they are getting more and more expansive each year. College education is about learning, which can be assisted by technology, but more importantly, it is also about student intellectual growth and maturing under the guidance of the nation's most highly educated population that loves life-long learning and share this experience with their students. Reality has proven that this group of highly educated people has been leading the renovation and innovation of our country.

The seventh and final theme was *Fearful and Anxious About Losing Their Employment*. For many faculty members the question of job security has become a new stressor in their lives, most have accepted the fact that their job will be different moving forward because of technology, but some are even questioning in this era of MOOC's and increasing online programs if their jobs will be needed moving forward. This question was advanced by Bower "As faculty members, we are warned that if we don't 'get with the program' our institutions will suffer and our jobs will be lost to more technologically, bottom-line oriented organizations such as the University of Phoenix" (p. 1) [53]. Faculty members resistant to technology and faced with a difficult and possibly frightening future, and we see in the following quote, "I think everything will be online and college classrooms will be a thing of the past - something that I find extremely scary."

Table 3. Faculty perceptions of the future of technology in higher education.

| # | Name of Theme | Explanation/Rule for Inclusion | Faculty Extracts |
|---|---|---|---|
| 1 | Increase in Technology Usage | Faculty predicted that future of education will see increased use of technology in higher education. | • I think there will be more and more technology … These types of technology were not around when I was growing up, but many of us (faculty) really enjoy learning the newer technologies and using them…(Bring on the wikis, blogs, podcasts, Youtube, etc.!!!)<br>• I think we will be forced to utilize more technology as the students become more dependent on it and less able to function without it. |
| 2 | Face-to-face Instruction and Physical Materials | Faculty predicted that the physical presence of classrooms and materials (i.e. textbooks) will continue into the future. | • Technology and face-to-face meetings should complement each other - neither should replace the other or be exclusive to the educational experience.<br>• I honestly hope that we continue to use technology to add to the educational experience, but that technology will not take the place of knowledgeable, caring faculty and the opportunity to interface person-to-person with faculty and students. |
| 3 | Hybrid Formats, Online Learning, | Faculty speculated that the future of education will be a varied format | • Dynamic interactive online education where faculty move away from being the Sage on the Stage to the facilitator and partner with students in teaching/learning. |
| | and Real-time Interaction | including online and hybrid and real-time interaction. | • Technology doesn't improve teaching, good teaching improves teaching. Technology is simply an additional tool in the toolbox. Used wisely, they support us. Current emphasis seems to simply be on encouraging wide-spread use without necessarily wide-spread need. |
| 4 | Technology Utilized to Prepare Students for Workplace | Faculty predicted that technology will be used to facilitate student preparation for career development. | • I think technology of the future will position students to function well in their chosen profession…<br>• We must recognize what technology is used in the work place and provide our students with the expertise they need for a seamless transfer. We have to realize that using "old" technolog[ies]…[are] still important in many workplaces and we do students a disservice when we don't insure they are savvy with these packages. |
| 5 | Online Learning | Faculty predicted that technology for education will only be available through online platforms. | • Probably simulations/virtual reality to replace what used to be hands-on experiences all in the name of saving money.<br>• Online instruction followed by personal and guided application.<br>• I think it will be primarily an online format. On the way to this, computer use in the classrooms will be increased. |
| 6 | The New Model of Education | Faculty predicted that technology will result in a new model of education for the future. | • I believe that the bricks and mortar universities will become more like cultural event venues and museums with all learning done online. The learning will be done in a much less structured environment where students select the information they want and access various sites throughout the world to take a class (in a MOOC-type format).<br>• The future of education is on the cusp of a quantum leap. With common courses being administered online to a mass market of students offered by Ivy League institutions, the transition has just begun. Frankly, I am unsure where we are headed, but we must be ready to adapt. |
| 7 | Fearful and Anxious About Losing Their Employment | Faculty are anxious that technology may replace them. | • Scary- you can overdo this.<br>• I think everything will be online and college classrooms will be a thing of the past - something that I find extremely scary. |

## 5. DISCUSSION AND CONCLUSION

### 5.1. Discussion

The success of any large scale programmatic shift towards course delivery methods which rely on greater use of technology is to a large extent dependent on the willingness and readiness of the faculty members to increase their use of technology in the classroom and online. The quantitative data in this project provide a snapshot of the current attitudes and skills of the faculty members as a whole. The modal response of faculty members in stating their preference for technology use was to use a "moderate" amount, indicating some reservation in having too much. However, there was considerable agreement with the statements indicating that in general, technology use enhances student engagement and learning, improves instruction and provides convenience in course delivery.

Reported use of technology in courses varied considerably, but indicated that on average, faculty members used approximately six different technologies such as word processing, spreadsheets, presentation software, course management systems, and the internet. They indicated their skill with such technologies as relatively high, though again there was considerable variability among instructors. The measure of how frequently they used those technologies again varied fairly widely, but was of moderate frequency across instructors.

As expected, the technology skill and usage variables were strongly correlated and were also related to respondents' attitudes toward technology use in courses. Also, those who favored extensive use of technology in the classroom had significantly higher levels of reported skill and frequency of use, and of a wider variety of technology tools, than did those favoring lesser amounts.

The study focused on assessing the faculty attitudes towards integrating technology into teaching. The main goal of the qualitative questions was to provide some insight into the actual use and opportunities, barriers, and the future direction of technology in the classroom. The qualitative findings support the quantitative data and the literature. However, some of the findings were inconsistent with the literature as indicated when faculty discussed that students many were not capable of using technology effectively [17][20].

The results of this study indicated technology provided both opportunities and challenges for faculty members as they attempted to use innovative classroom practices within higher education. The technologies that enabled classroom instruction included a variety of tools, applications, and resources. Faculty members noted that technology facilitated the organization of their courses, enabled student collaboration, and deterred plagiarism. According to the USDE, "Technology infuses classrooms with digital learning tools…expands course offerings, experiences, and learning materials; supports learning 24 hours a day, 7 days a week; builds 21[st] century skills; increases student engagement and motivation; and accelerates learning" (para. 1) [35]. Although there were numerous opportunities for technology use, one faculty member pointed out, "No single technology is more useful than another, because they each contribute to a rich learning experience for the student."

The faculty members reported that they had numerous barriers to success with technology. One such barrier was their own lack of knowledge. A compelling barrier was that of a lack of knowledge of technology skills by students. Kvavik noted that current students need to develop both information literacy as well as the technical skills for utilizing technology tools in the classroom and remarked that they "tend to know just enough technology functionality to accomplish their work; they have less in-depth application knowledge or problem solving skills" (para. 12) [49].

The future of technology is promising, and our findings are consistent and inconsistent with the literature. For example, faculty members report embracing technology, but they also are fearful that the high-quality interaction of the face-to-face classroom will vanish as indicated by a faculty member's comment regarding the loss of the humanistic perspective in education. Faculty members saw technology as a way to make an emotional connection. This is illustrated in the comment: "High tech, high touch. The technology that can touch students emotionally will win." Other ideas such as the concern that the future of technology is so fluid that it is hard to predict or that there will be less interaction in the learning environment and more isolation. Mobile learning is also an emerging theme as faculty members speculated that technology would become mobile, for anytime, anywhere learning. The increasing number of smartphones and tablets is fuelling this idea.

One faculty member reported that they see their role as an orchestrator, and this reflects the direction of technology forecasted in the literature [17]. Innovation was reflected in the literature review, and this was a consistent theme among faculty members' responses. It is important that as faculty members think about the future of technology in the classroom, the university and administration provide faculty members with the tools and rewards to use technology to teach.

## 5.2. Limitations

This research contributes to the knowledge of faculty members' attitudes towards technology in the classroom. The study focused on faculty members' attitudes regarding opportunities, barriers, and the future of technology, but data was self-reported and anonymous and therefore unverifiable. The results presented in the research have limited external validity as a result of using a non-validated survey. Additionally, the survey was written to assess the needs of the particular institution's faculty, and it may not be generalizable to other university settings. Further, the sample in this research study is small, and there were disciplines under-represented in the study.

## 5.3. Conclusions and Future Direction

The future of higher education will progressively become more focused on increasing access to quality education opportunities, improved degree completion rates, and containing costs to both the university and the student. President Obama's goal to raise the Nation's college graduation completion rate to 60% by 2020 is a clear indication of the future direction of the U.S. Higher Education System [54]. Moreover, the President's most recent plan to begin rating universities based on performance and value will continue to drive creative and innovative models of higher education. To achieve the aforementioned goals, universities will need to engage in innovation and competition to maintain high quality educational experiences, while at the same time increasing accessibility, affordability and graduation rates.

To meet the challenges of the 21st Century, the university highlighted in this case has redesigned 10 of its most popular majors to increase student success, flexibility of scheduling, accessibility, and graduation rates. This five-year plan will provide a hybrid option for any student participating in the program. One researcher analyzed many hybrid redesign projects and found that institutions could utilize technology to reduce the number of class meetings on campus, thereby increasing the accessibility [55]. Furthermore, these redesign projects resulted in significantly lower number of students receiving either non-passing grades or withdrawing from the course. With a reduced number of students having to retake their courses, the universities increased retention rates and improved time to degree completion.

While the results of current study provide evidence that faculty from the university are open to the use of technology, the success of any technology rich program is rooted within the preparedness and willingness of the faculty and students to embrace this new teaching and learning paradigm. Therefore, to ensure the success of the aforementioned redesign project, the university has made a commitment to both faculty and student development as it relates to the use of technology enriched teaching and learning environments.  Furthermore, as the five-year plan moves forward, extensive data will be collected to determine the impact of this major programmatic shift on student learning outcomes, faculty and student satisfaction and cost savings for both the university and student.

## Authors

**Colleen Marzilli** is an Assistant Professor in the College of Nursing and Health Sciences at The University of Texas at Tyler. She is a PhD candidate, and she has a Doctorate of Nursing Practice with an emphasis in Public Health Nursing from The University of Tennessee Health Science Center at Memphis. Her areas of focus include best practices in teaching, cultural competence within the education and health care settings, and cultural implications related to health disparities and health systems. Colleen has worked in the health care setting for over eleven years in various roles, and she specialty certifications in medical-surgical nursing, case management, public health nursing, and nursing education. 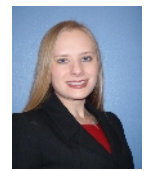

**Julie Delello** is an Assistant Professor in the College of Education and Psychology at The University of Texas at Tyler. She received her PhD in Curriculum and Instruction with a specialization in science and technology from Texas A&M University. Her areas of focus include Response to Intervention, Disability Studies, Visual Media Technologies, Virtual Science Museums, Social Media Pl atforms, and ePortfolios for authentic learning. Julie has worked in K-12 education for over 20 years as both a teacher and as an administrator. Julie helped to design virtual science museums in conjunction with The Chinese Academy of Sciences, Computer Network Information Center in Beijing, China. In addition, Julie has won several grants and teaching awards including a National Science Foundation Grant for The East Asia and Pacific Summer Institutes, the Golden Apple Educator Award for Best Practices in Staff Development and Curriculum Initiatives, and the 2012 University of Texas at Tyler-Kappa Delta Pi Teacher of the Year award. Julie was also the invited guest speaker at the United States Department of State Eleventh Annual Joint U.S.-China Joint Science and Technology Commission Meeting on the efforts of expanding the scientific and educational ties between the U.S. and China. 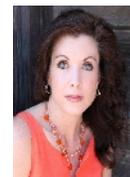

**Shelly Marmion** is a Professor and Associate Chair of Undergraduate Programs in the Department of Psychology and Counseling at The University of Texas at Tyler. She received her PhD in Experimental Psychology with concentrations in Cognitive Psychology and Statistics from Texas Tech University. She has over 35 years of teaching experience in higher education. Shelly has worked as a statistical and design consultant for various funding agencies, research center s and community organizations, and authored numerous articles and a book on various vulnerable populations. She has also received numerous teaching awards and is one of 26 inaugural faculty members in the PATSS program where she has created innovative ways to use technology in the classroom and engage her students. 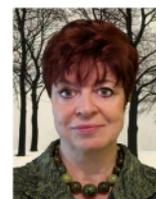

**Rochell R. McWhorter** is an Assistant Professor of Human Resource Development in the College of Business and Technology at The University of Texas at Tyler. She received her Ph.D. degree in Human Resource Development from Texas A&M University. She has over 20 years' experience in industry and K-12 education. Rochell has edited and authored a number of journal articles and scholarly resources on technology-facilitated learning in higher education. Her scholarly publications include topics such as ePortfolios as facilitators of learning and professional branding, virtual human resource development, visual social media, scenario planning for leadership development, and virtual learning environments for real-time collaboration. She has been a of numerous teaching awards including the Silvius-Wolansky Outstanding Young Teacher Educator Award from the Association for Career and Technical Education. She serves on numerous committees promoting the effective use of technology in higher education and Chairs the Virtual HRD, Technology, and e-Learning Special Interest Group for the Academy of Human Resource Development. 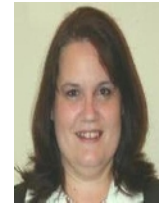

**Paul B. Roberts** is currently the Associate Dean in the College of Business and Technology and Associate Professor in the Department of Human Resource Development and Technology at The University of Texas at Tyler. He has received over $900,000 in grants and funded projects since coming to UT Tyler in 1992. He is the director for an online teacher training grant with the Texas Education Agency and coordinates ongoing professional development activities. His research focuses on Virt ual HRD and the demographics of HRD programs. P aul serves as editor of the "Directory of HRD Programs in the US". He has received numerous honors and awards for teaching including recently awarded the Chancellor's Council Outstanding Teaching Award and nominated for the Regents' Outstanding Teaching Award. Paul earned his doctorate with an emphasis in computer assisted instruction and curriculum development from the Department of Human Resource Development at Texas A&M University in 1994. 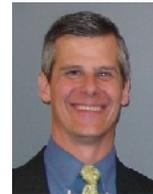

**T. Scott Marzilli** is an Assistant Vice President for Teaching Innovation and Student Success at The University of Texas at Tyler. His responsibilities include Online Education, Center for Teaching Excellence and Innovation, and the Department of Academic Success. Currently he is coordinating the Patriots Applying Technology for Success and Savings (PATSS) initiative at The University of Texas at Tyler. He formally was the Interim Dean for the College of Nursing and Health Sciences and the Department Chair for the Health and Kinesiology. 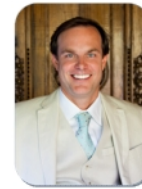